%% file: main.tex
\documentclass[conference]{IEEEtran}

\usepackage[utf8]{inputenc}

\usepackage{spconf,amsmath,amssymb,graphicx}
\usepackage{bm}
\usepackage{color}
\usepackage{algpseudocode}
\usepackage{algorithm}
\DeclareMathOperator*{\argmax}{arg\,max}
\DeclareMathOperator*{\argmin}{arg\,min}

\usepackage{subcaption}
\usepackage{pgfplots}
\usepackage{tikz}
\usepackage{caption}
\usepackage{siunitx}
\usetikzlibrary{patterns,shapes.arrows}
\pgfplotsset{compat=newest}
\pgfplotsset{every tick label/.append style={font=\footnotesize}}

\title{Riemannian classification of EEG signals with missing values}
\name{A. Hippert-Ferrer$^1$, A. Mian$^2$, F. Bouchard$^1$, F. Pascal$^1$\thanks{This work was supported by ANR-ASTRID MARGARITA under Grant ANR-17-ASTR-0015.}}
\address{
	$^1$ Universit\'e Paris-Saclay, CNRS, CentraleSup\'elec, Laboratoire des signaux et syst\`emes, \\
	91190, Gif-sur-Yvette, France\\
	$^2$ LISTIC, Université Savoie Mont Blanc 
}
\begin{document}
%
\maketitle
\begin{abstract}
	This paper proposes a strategy to handle missing data for the classification of electroencephalograms using covariance matrices. It relies on the observed-data likelihood within an expectation-maximization algorithm. This approach is compared to two existing state-of-the-art methods: \emph{(i)} covariance matrices computed with imputed data; \emph{(ii)} Riemannian averages of partially observed covariance matrix. All approaches are combined with the minimum distance to Riemannian mean classifier and applied to a classification task of two widely known paradigms of brain-computer interfaces. In addition to be applicable for a wider range of missing data scenarios, the proposed strategy generally performs better than other methods on the considered real EEG data.
\end{abstract}

\section{Introduction}
\label{sec:intro}

Electroencephalography (EEG) is a non-invasive neuroimaging modality pioneered by Hans Berger~\cite{B29}, which consists in recording the electrical activity of the brain by placing electrodes on the scalp.
The low cost, simplicity and high temporal resolution (it captures well the brain activity dynamics) of EEG have made its popularity and allowed for its use in various applications.
EEG is indeed widely used in brain research for instance to study sleep or epilepsy.
It is also central for brain-computer interfaces (BCI), where the subject interacts with a computer through brain signals.
These can for example be employed to control exoskeleton~\cite{KCBDMH16} or help mechanical ventilation~\cite{CBHMMALA18}.

This paper focuses onto BCI and more specifically onto two of the various possible paradigms: event related potentials (ERP) and Motor Imagery (MI). ERP are specific time-locked responses of the brain to some stimuli, such as flashes~\cite{wolpaw2002}. Given the brain signals, the goal is to determine if the subject was exposed to the stimulus, \emph{i.e.}, find out if the specific time-locked response is in the EEG recording. 
Concerning MI, the goal is to recognize a mental task (as hand or foot movement) which is represented in different areas of the motor cortex.
In both cases, BCI boils down to a classification task of brain signals at hand.
Various methods have specifically been designed to achieve this; see~\cite{LBCCCRY18} for a recent review.
In this work, we consider one of the most efficient approach, called minimum distance to Riemannian mean (MDRM)~\cite{BBCJ11}, which exploits covariance matrices of the data along with their natural Riemannian geometry.

BCI classification methods have shown good performance on a number of classical datasets; see \emph{e.g.},~\cite{jayaram2018}.
However, these datasets have been recorded in laboratories, which are well controlled environments, and data have been curated by hand to ensure the quality of the recordings.
In real life scenarios, obtaining such idealistic data seems unrealistic.
One can expect the data at hand to contain faulty measurements due for instance to impedence change, electrode disconnections~\cite{bahador2020} or artifacts coming from movements of the subject, blinks, \emph{etc.}
These can lead to missing data in EEG recordings as one may decide to discard it from the data.
When it comes to classification tasks, missing data impedes the exploitation of complementary information in non-missing channels.
Considering this, it is of primary importance to develop an appropriate procedure to classify incomplete EEG signals.

To achieve this, several strategies can be employed.
A first possibility is to exploit some missing data imputation technique in order to retrieve a complete EEG recording.
It has for instance been considered for EEG in~\cite{perrin1989} with an algorithm relying on spherical spline interpolation.
Imputation can also be performed through the $K$-nearest neighbors (KNN) algorithm~\cite{troyanska2001}.
Another possibility is to directly work in the covariance space.
When a channel is missing, an imputation of symmetric positive definite (SPD) matrices with uncorrelated white noise has been proposed in~\cite{rodrigues2019}. In a similar vein, the authors of \cite{yger2020} propose to estimate the Riemannian mean of partially observed covariance matrices with missing variables (channels) to later perform a classification task with the MDRM.
When the data contains missing samples\footnote{To appreciate the different nature of works handling missing samples in the data and those treating on missing variables in the data and/or the covariance matrix, see Table~1 in \cite{yger2020}.}, it is possible to estimate the covariance matrix of incomplete data using an expectation-maximization (EM) algorithm~\cite{dempster1977}.
More recently, a neural network approach was designed to interpolate missing electrodes in EEG \cite{saba2020}.

In this paper, we propose to adapt the state-of-the-art EEG classification pipeline~\cite{BBCJ11} to handle missing data by exploiting an EM algorithm to compute covariance matrices.
Compared to two state-of-the-art approaches -- missing data imputation using KNN~\cite{troyanska2001} and Riemannian averaging over incomplete covariance matrices~\cite{yger2020} -- our method is applicable in more missing data scenarios.
Furthermore, experiments conducted onto real ERP and motor imagery~(MI) EEG data tend to show that our method is competitive in terms of performance as compared to others.

\section{EEG classification for complete data}
\label{sec:classif_classical}

To classify EEG data, rather than directly using the raw data, one usually exploits their covariance matrices as features~\cite{BBCJ11}.
The classification of these covariance matrices is then performed by leveraging their Riemannian geometry: a well-adapted distance is used to compute centers of mass of the different classes and assigning the class to unknown recordings.

\subsection{Covariance matrix estimation}

A common hypothesis for EEG signals is to assume a multivariate normal distribution \cite{congedo2008}. Let $\bm{X} = [\bm{x}_1,\dots,\bm{x}_n] \in \mathbb{R}^{p\times n}$ be a recorded EEG trial containing $p$-dimensional real vectors, where $p$ is the number of electrodes and $n$ the time samples. $\bm{X}$ follows a zero-mean Gaussian distribution with covariance matrix $\bm{\Sigma} \in \mathbb{R}^{p\times p}$, denoted $\bm{X} \sim \mathcal{N}(\bm{0}, \bm{\Sigma})$. When the data is complete, the Maximum Likehood Estimator (MLE) is the well-known Sample Covariance Matrix (SCM):

\begin{equation}	
\label{eq:SCM}
\bm{\Sigma} = \frac{1}{n}\bm{X}\bm{X}^\top
\end{equation}

\subsection{Minimum distance to Riemannian mean}

The idea consists in leveraging a non-Euclidean distance based on the geometry of the space of SPD matrices (thus covariance matrices). In peculiar, theoretical studies have shown benefits to use the so-called affine invariant distance~\cite{pennec2020}. Given two covariances matrices $\boldsymbol{\Sigma}_1, \boldsymbol{\Sigma}_2$, its expression is given by:
\begin{equation}
\delta_\mathcal{AF}(\boldsymbol{\Sigma}_1, \boldsymbol{\Sigma}_2) = \left\|\log \left(\boldsymbol{\Sigma}_1^{-1 / 2} \boldsymbol{\Sigma}_2 \boldsymbol{\Sigma}_1^{-1 / 2}\right)\right\|_{\mathrm{F}},
\end{equation}
where $\|\cdot\|_{\mathrm{F}}$ is the Frobenius norm and $\log$ is the matrix logarithm.
Then, given a classification problem of $Z$ classes, with training data $\{(\boldsymbol{\Sigma}_{\ell}, y_{\ell}):1\leq \ell\leq L\}$, the training phase consists in computing the mean of each class $\{\overline{\boldsymbol{\Sigma}}^{(z)}:z\in\{1,\dots,Z\}\}$ according to the affine-invariant distance by using the iterative algorithm of~\cite{jeuris2012}. A predicted label $y$ is then obtained on a new sample $\boldsymbol{\Sigma}$ by assigning it to the closest mean:
\begin{equation}
\label{eq:label_pred}
y = \underset{z\in\{1,\dots,Z\}}\argmin\delta_\mathcal{AF}(\boldsymbol{\Sigma}, \overline{\boldsymbol{\Sigma}}^{(z)})
\end{equation}

\section{Existing strategies to handle missing data for EEG classification}
\label{sec:missing_classical}
\subsection{Data imputation: KNN algorithm}
\label{subsec:imputation}
A first possibility to handle missing data is to impute them, which then allows to compute their covariance matrix according to~\eqref{eq:SCM} -- and thus apply the MDRM classification method.
We focus here on $K$-nearest neighbors (KNN) interpolation \cite{troyanska2001}.
If the $j^{\text{th}}$ electrode of $\bm{x}_i$ is missing, KNN finds its $K$-closest neighbors in the training set according to an Euclidean distance $d\big(\bm{x}_i,\{\bm{x}_q\}_{q=1,q\ne i}^n\big)$ compatible with missing data, called the heterogeneous Euclidean-overlap metric (HEOM)~\cite{garcia2009}. If $\{\bm{v}_k\}_{k=1}^K$ represents the set of $K$-selected neighbors, then the imputed value $\hat{x}_{ij}$ is

\begin{equation}
\hat{x}_{ij} = \frac{1}{K}\sum_{k=1}^{K}w_k v_{kj}, \;\; w_k = \frac{1}{d(\bm{x}_i, \bm{v}_k)^2}, \;\; \sum_{k=1}^{K}w_k=1
\end{equation}
where $w_k$ denotes the corresponding weight of the $k$-th nearest neighbor $\bm{v}_k$ and $d(\bm{x}_i, \bm{v}_k)^2$ is the squared HEOM. The choice of $K$ can be determined by cross-validation on observed data only, which can be challenging with high missing data ratio.

\subsection{Masked Minimum to Riemannian mean}

In \cite{yger2020}, authors propose an implementation of the MDM algorithm for incomplete data. Such design requires to construct a mask to extract\footnote{Such matrix is often called selection matrix.} the observed data from an incomplete input $\bm{X} \in \mathbb{R}^{p\times n}$. A mask $\bm{M} \in \mathbb{R}^{p\times (p-r)}$ is defined as the identity matrix with columns of the $r$ missing variables removed.  If $\bm{X}$ is incomplete, a complete $(p-r)\times n$ submatrix $\check{\bm{X}}$ can be extracted using $\bm{M}^\top\bm{X}$. The covariance matrix becomes

\begin{equation}
\check{\bm{\Sigma}} = \frac{1}{n}\bm{M}^\top\bm{\Sigma}\bm{M}
\end{equation} 
where $\bm{\Sigma}$ is the complete covariance matrix as defined in~(\ref{eq:SCM}). The next step is to consider the set of $\{\bm{\Sigma}_i\}_{i=1}^n$ incomplete SPD matrices, where each $\bm{\Sigma}_i$ has $r_i$ missing variables and an associated mask $\bm{M}_i$. For each class $z\in\{1,\dots,Z\}$, the masked Riemannian mean $\bar{\bm{\Sigma}} \in \mathcal{S}_{++}^p$ is defined as:
\begin{equation}
	\overline{\bm{\Sigma}}^{(z)} = \argmin\limits_{\boldsymbol{\Sigma}\in\mathcal{S}_{++}^p} \frac{1}{2} \sum_{i=1}^{n} w_i \delta^2_\mathcal{AF}(\bm{M}_i^\top \bm{\Sigma}_i \bm{M}_i,\bm{M}_i^\top \bm{\Sigma} \bm{M}_i)
\end{equation}
Finally, the label prediction is performed as in~(\ref{eq:label_pred}) with the suited mask added.

\section{An EM algorithm for covariance estimation with missing data}
\label{sec:missing_em}

A powerful strategy to deal with missing values is to directly estimate the covariance feature without imputation. In this case, one can rely on maximum likelihood estimation using the EM algorithm developed in \cite{hippertferrer2021}. Here, the unknown parameter to estimate is simply $\bm{\theta} = \{\bm{\Sigma}\}$. In the following EM algorithm, $\bm{x}_i$ is transformed into $\widetilde{\bm x}_i$ such that $\widetilde{\bm x}_i = \bm P_i \bm x_i = \begin{pmatrix}
\bm{x}_{i}^\text{o} & \bm{x}_{i}^\text{m}
\end{pmatrix}^\top$, where $\bm{P}_i \in \mathbb{R}^{p\times p}$ is a permutation matrix, $\bm x_i^\text{o}$ and $\bm x_i^\text{m}$ denote the observed and missing elements of $\bm{x}_i$, respectively. The covariance matrix of $\widetilde{\bm x}_i$ is

\begin{equation}
\widetilde{\bm{\Sigma}}_{i}=\begin{pmatrix}
\widetilde{\bm{\Sigma}}_{i,\text{oo}} & \widetilde{\bm{\Sigma}}_{i,\text{mo}} \\
\widetilde{\bm{\Sigma}}_{i,\text{om}} & \widetilde{\bm{\Sigma}}_{i,\text{mm}}
\end{pmatrix}=\bm{P}_i\bm{\Sigma}\bm{P}_i^{\top}
\end{equation}
where $\widetilde{\bm{\Sigma}}_{i,\text{mm}}$, $\widetilde{\bm{\Sigma}}_{i,\text{mo}}$, $\widetilde{\bm{\Sigma}}_{i,\text{oo}}$ are the block CM of $\bm{x}_i^\text{m}$, of $\bm{x}_i^\text{m}$ and $\bm{x}_i^\text{o}$, and of $\bm{x}_i^\text{o}$.
Replacing $\bm{x}_i$ by its permuted version $\widetilde{\bm{x}}_i$, the \textit{complete} log-likelihood for the Gaussian distribution is expressed as: 

\begin{align}
\mathcal{L}_c(\bm{\theta}|\bm{x}) &\propto -n\log|\bm{\Sigma}| - \sum_{i=1}^{n}\begin{pmatrix}
\bm{x}_{i}^\text{o} \\
\bm{x}_{i}^\text{m}
\end{pmatrix}^\top \widetilde{\bm{\Sigma}}_i^{-1} \begin{pmatrix}
\bm{x}_{i}^\text{o} \\
\bm{x}_{i}^\text{m}
\end{pmatrix}
\end{align}
The EM algorithm, which is given in Algorithm~\ref{alg:sigma}, can now be formulated.

\noindent
\textbf{Initialization}. At step $t=0$ of the algorithm, initialize the covariance matrix with the SCM computed from fully observed $\bm{x}_i$ (\textit{i.e.}, no missing values), denoted $\bm{\Sigma}_{\text{SCM-obs}}$.

\noindent
\textbf{E-step}. Compute the expectation of the missing data conditioned by the observed data and the estimated covariance at the $t$-th iteration:

\begin{align}
Q_i(\bm{\theta}|\bm{\theta}^{(t)}) &= \mathbb{E}_{\bm{x}_i^\text{m}|\bm{x}_i^\text{o},\bm{\theta}^{(t)}} \big[\mathcal{L}_c(\bm{\theta}|\bm{x}_i^\text{o},\bm{x}_i^\text{m})\big] \nonumber \\
&= \text{tr}(\bm{B}_i^{(t)}\widetilde{\bm{\Sigma}}_i^{-1})
\end{align}
with $\bm{B}_i^{(t)} = \begin{pmatrix}
\bm{x}_i^\text{o} \bm{x}_i^{\text{o}^\top} & \mathbb{E}_{\bm{x}_i^\text{m}|\bm{x}_i^\text{o},\bm{\theta}^{(t)}}[\bm{x}^\text{o}_i\bm{x}^{\text{m}^\top}_i] \\
\mathbb{E}_{\bm{x}_i^\text{m}|\bm{x}_i^\text{o},\bm{\theta}^{(t)}}[\bm{x}^{\text{m}}_i\bm{x}^{\text{o}^\top}_i] & \mathbb{E}_{\bm{x}_i^\text{m}|\bm{x}_i^\text{o},\bm{\theta}^{(t)}}[\bm{x}^\text{m}_i\bm{x}^{\text{m}^\top}_i]
\end{pmatrix}$. The only expectation to compute are $\mathbb{E}_{\bm{x}_i^\text{m}|\bm{x}_i^\text{o},\bm{\theta}^{(t)}}[\bm{x}^\text{m}_i]$ and $\mathbb{E}_{\bm{x}_i^\text{m}|\bm{x}_i^\text{o},\bm{\theta}^{(t)}}[\bm{x}^\text{m}_i\bm{x}^{\text{m}^\top}_i]$, which are the expectation of the sufficient statistics $\bm{x}^\text{m}_i$ and $\bm{x}^\text{m}_i\bm{x}^{\text{m}^\top}_i$ (denoted $\mathbb{E}\big[\bm{x}^\text{m}_i\big]$ and $\mathbb{E}\big[\bm{x}^\text{m}_i\bm{x}^{\text{m}^\top}_i\big]$ hereafter). A classical result in conditional distributions \cite{anderson1965} gives the following expressions:

\begin{align}
\mathbb{E}\big[\bm{x}^\text{m}_i\big] &= \widetilde{\bm{\Sigma}}_{i,\text{mo}}\widetilde{\bm{\Sigma}}_{i,\text{oo}}^{-1}\widetilde{\bm{x}}_i^\text{o} \\
\mathbb{E}\big[\bm{x}^\text{m}_i\bm{x}^{\text{m}^\top}_i\big] &= \widetilde{\bm{\Sigma}}_{i,\text{mm}} - \widetilde{\bm{\Sigma}}_{i,\text{mo}}\widetilde{\bm{\Sigma}}_{i,\text{oo}}^{-1}\widetilde{\bm{\Sigma}}_{i,\text{om}} + \mathbb{E}\big[\bm{x}^\text{m}_i\big]\mathbb{E}\big[\bm{x}^\text{m}_i\big]^\top
\end{align}

\noindent
\textbf{M-step}. Resolve the following maximization problem:

\begin{equation}
\label{eq:argmaxQ}
\bm{\Sigma}^{(t+1)} = \argmax\limits_{\boldsymbol{\Sigma}\in\mathcal{S}_{++}^p} \sum_{i=1}^{n}Q_i(\bm{\Sigma}|\bm{\Sigma}^{(t)})
\end{equation}
where $\mathcal{S}_{++}^p$ is the set of symmetric positive definite (SPD) matrices. Solving (\ref{eq:argmaxQ}) requires to solve $\frac{\partial Q_i}{\partial \bm{\Sigma}}=0$. Fortunately in this case, the following closed-form expression can be found with simple derivation calculus:

\begin{equation}
\bm{\Sigma}^{(t+1)} = \frac{1}{n}\sum_{i=1}^n \bm{P}_i\bm{B}_i^{(t)\top}\bm{P}_i^\top
\end{equation}

\begin{algorithm}
	\begin{algorithmic}[1]
		\Require $\{\widetilde{\bm{x}}_i\}_{i=1}^n \sim \mathcal{N}(\bm{0}, \bm{\Sigma}), \{\bm{P}_i\}_{i=1}^n$
		\Ensure $\widehat{\bm{\Sigma}}$
		\State Initialization: $\bm{\Sigma}^{(0)} = \widehat{\bm{\Sigma}}_{\text{SCM-obs}}$
		\Repeat 
		\State Compute 	$\bm{B}_i^{(t)} = \begin{pmatrix}
		\bm{x}_{i}^\text{o}\bm{x}_i^{o\top} & \mathbb{E}[\bm{x}^\text{o}_i\bm{x}^{\text{m}^\top}_i] \\
		\mathbb{E}[\bm{x}^\text{m}_i\bm{x}^{\text{o}^\top}_i] & \mathbb{E}[\bm{x}_i^\text{m}\bm{x}_i^{\text{m}\top}]
		\end{pmatrix}$
		\State Compute $\bm{\Sigma}^{(t+1)} = \frac{1}{n}\sum_{i=1}^n \bm{P}_i\bm{B}_i^{(t)\top}\bm{P}_i^\top$
		\State $t \leftarrow t+1$
		\Until $||\bm{\Sigma}^{(t+1)} - \bm{\Sigma}^{(t)}||^2_F$ converges
		
	\end{algorithmic}
	\caption{Estimation of $\bm{\Sigma}$ with the EM algorithm}
	\label{alg:sigma}
\end{algorithm}

\section{Numerical experiments}
\label{sec:num_exp}

\subsection{Data}

The proposed procedures are assessed through a detection task of EEG recordings of two distinct BCI experiments\footnote{Both data sets are accessed through the \texttt{moabb} database~\cite{jayaram2018}.}:
\begin{itemize}
	\item[i)] \textit{P300 Event Related Potentials} triggered from target stimuli consisting of flashes~\cite{wolpaw2002}. This dataset consists in 3 recording sessions of EEG potentials on 10 subjects using 16 electrodes\footnote{Fz, FCz, Cz, CPz, Pz, Oz, F3, F4, C3, C4, CP3, CP4, P3, P4, PO7, PO8. Please refer to \cite{aric2014} for an extensive description of the dataset.}. The signals are band-pass filtered between 1 and 20Hz,  downsampled to 128Hz and epoched into 1728 trials: one trial out of six corresponds to the flash stimulus (class 1) while the rest correspond to the absence of target (class 0).
	Before computing covariance matrices, raw data are augmented by concatenating the prototyped ERP response as prescribed in~\cite{barachant2014}.
	This leads for each subjects to a dataset of $L=1728$ covariance matrices as in~\eqref{eq:SCM} with $p=16$ and $n=103$.
	
	\item[ii)] \textit{Motor Imagery}: this dataset \cite{tangermann2012} consists of EEG data from 9 subjects recorded using 22 electrodes. This BCI paradigm consists of four different motor imagery tasks, namely the imagination of movement of the left hand (class 1), right hand (class 2), both feet (class 3), and tongue (class 4). Two sessions were recorded for each subject, with a total of 288 trials per session. This leads for each subjects to a dataset of $L=576$ covariance matrices with $p=22$ and $n=1001$.

\end{itemize}

\subsection{Missing data scenarios}

\begin{figure}
	\centering
	\input{./img/dessin.tex}
	\caption{Illustration of electrode popping (red) and eye blinking (purple).}
	\label{fig:1}
\end{figure}
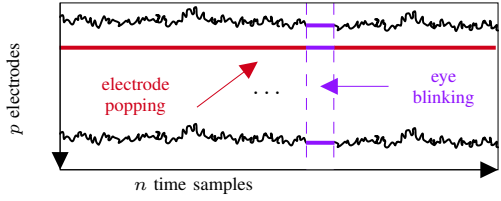

We consider two realistic scenarios of incomplete EEG signals: \textit{electrode popping} (s1) and \textit{eye blinking} (s2). As illustrated in~Fig.\ref{fig:1}, electrode popping yields one or multiple missing electrodes during signal acquisition. Eye blinking creates an undesired artifact over a short time period (typically $\sim$200 ms) over a group of electrodes. Such artifacts are then removed, which creates missing data.
As shown in Table~\ref{tab:methods_comp}, compared to the other methods considered, our proposed strategy is the only one working in both cases.

\begin{table}
	\begin{tabular}{c|c|c|c}
		& KNN \cite{troyanska2001} & Masked mean \cite{yger2020} & EM \cite{hippertferrer2021} \\
		Electrode popping (s1)& $\times$ & \checkmark & \checkmark \\
		Eye blinking (s2) & \checkmark & $\times$ & \checkmark
	\end{tabular}
	\caption{Applicability (\checkmark: applicable; $\times$: not applicable) of the proposed methods for different missingness scenarios.}
	\label{tab:methods_comp}
\end{table}

For scenario s1, two electrodes are entirely removed. For scenario s2, three groups of electrodes (11 for P300 and 12 for MI) during three time intervals are removed. In both cases, removals are done on an increasing number of trials to appreciate the effect on classification accuracy.

\subsection{Results}

To compare the accuracy of the classification of P300 and Motor Imagery datasets, the MRDM classifier as described in section~\ref{sec:classif_classical} is used. The following strategies are considered to estimate the covariance matrices: \underline{\textbf{SCM}}: the SCM of \textit{complete} data, \textit{i.e.}, without missing values;
\underline{\textbf{KNN + SCM}}: KNN imputer implemented from the \texttt{scikit-learn} package~\cite{pedregosa2011} applied to incomplete data with a number of $k=5$ neighbors.
\underline{\textbf{Masked-SCM}}: MDRM classifier using a masked version of the Riemannian mean~\cite{yger2020}.
\underline{\textbf{EM-SCM}}: hybrid strategy where the covariance matrix is estimated through Algorithm~\ref{alg:sigma} and fed into the MDRM classifier.

Performance is evaluated using the classical training-test paradigm with stratified $K$-folding ($K=5$). Algorithms are calibrated on data recorded during the training phase, and applied on data recorded during the test phase.

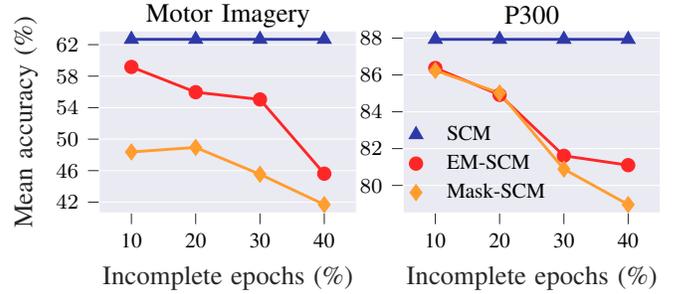
\begin{figure}
	\centering
	\input{./img/popping.tex}
	\caption{\underline{Electrode popping experiment}: classification mean accuracy as function of the incomplete epochs ratio.}
	\label{fig:2}
\end{figure}

\noindent
\textbf{Electrode popping experiment.} Accuracy results for this experiment are presented in Figure~\ref{fig:2}. The best performance is achieved on complete data (SCM), which serves as a baseline.
EM-based classification achieves equal or higher scores than the masked MDRM for both datasets.
On these data, our proposed strategy thus appears more advantageous than the masked version of the Riemannian mean.

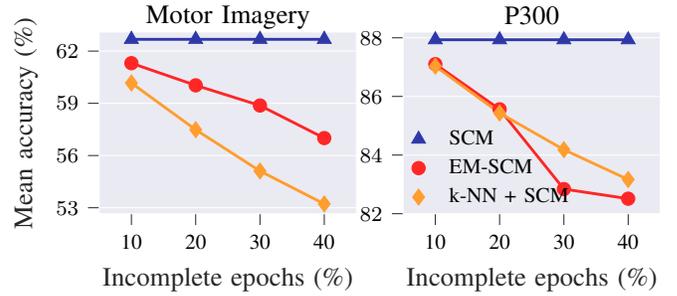
\begin{figure}
	\centering
	\input{./img/eye_blinking.tex}
	\caption{\underline{Eye blinking experiment}: classification mean accuracy as function of the incomplete epochs ratio.}
	\label{fig:3}
\end{figure}

\noindent
\textbf{Eye blinking experiment.} Results for this experiment are presented in Figure~\ref{fig:3}.
For the motor imagery dataset, KNN imputation does not perform as well as the EM-based classification. For the P300 dataset, KNN imputation shows higher accuracies as compared to the EM algorithm when the amount of missing data increases.

\section{Conclusion}

In this paper, a new strategy to handle missing data in the context of EEG classification is proposed.
It relies on an EM algorithm to directly estimate covariance matrices in the presence of missing data.
As compared to two other state-of-the-art strategies -- based on KNN imputation and masked Riemannian means -- our proposed method works with a wider range of missing data scenarios.
In addition, experiments conducted on real EEG data tends to show that our method is also competitve in terms of accuracy results.


\bibliographystyle{IEEEtran}
\bibliography{refs}

\end{document}

%% file: img/dessin.tex
\tikzset{every picture/.style={line width=0.35pt}} 

\begin{tikzpicture}[x=0.75pt,y=0.75pt,yscale=-1,xscale=1]

\draw [color={rgb, 255:red, 208; green, 2; blue, 27 }  ,draw opacity=1 ][line width=1.5]    (30,23) -- (250,23) ;
\draw    (30.28,84.61) -- (248.15,84.61) ;
\draw [shift={(251.15,84.61)}, rotate = 180] [fill={rgb, 255:red, 0; green, 0; blue, 0 }  ][line width=0.08]  [draw opacity=0] (8.93,-4.29) -- (0,0) -- (8.93,4.29) -- cycle    ;
\draw    (30.28,81.61) -- (30.28,0.39) ;
\draw [shift={(30.28,84.61)}, rotate = 270] [fill={rgb, 255:red, 0; green, 0; blue, 0 }  ][line width=0.08]  [draw opacity=0] (8.93,-4.29) -- (0,0) -- (8.93,4.29) -- cycle    ;
\draw  [color={rgb, 255:red, 0; green, 0; blue, 0 }  ][line width=0.75] [line join = round][line cap = round] (30.28,7.54) .. controls (30.28,7.14) and (31.43,6.26) .. (31.65,6.63) .. controls (32.32,7.73) and (31.86,8.96) .. (32.2,10.1) .. controls (32.36,10.6) and (33.04,8.67) .. (33.86,8.89) .. controls (35.25,9.28) and (35.1,10.49) .. (36.06,11.16) .. controls (37.1,11.89) and (37.35,9.45) .. (37.99,8.59) .. controls (38.65,7.72) and (39.24,9.31) .. (39.37,9.05) .. controls (39.7,8.31) and (38.93,7.32) .. (39.92,6.78) .. controls (41.54,5.89) and (40.58,8.08) .. (41.02,8.44) .. controls (41.6,8.92) and (42.68,7.04) .. (42.95,7.54) .. controls (43.44,8.44) and (43.33,9.49) .. (44.32,10.25) .. controls (45.17,10.9) and (45.6,7.19) .. (45.7,7.99) .. controls (45.87,9.3) and (45.77,10.66) .. (46.53,11.91) .. controls (47.04,12.76) and (47.47,8.49) .. (47.9,9.35) .. controls (48.4,10.34) and (48.17,11.45) .. (49,12.37) .. controls (49.21,12.59) and (49.7,13.21) .. (49.83,12.97) .. controls (50.83,11.08) and (51.31,9.12) .. (52.31,7.23) .. controls (52.68,6.53) and (53.37,9.14) .. (53.41,9.2) .. controls (53.65,9.52) and (54.43,7.63) .. (55.06,8.59) .. controls (55.5,9.26) and (54.07,10.12) .. (54.79,10.71) .. controls (55.58,11.36) and (56.53,8.39) .. (56.99,8.89) .. controls (57.65,9.62) and (59.09,9.69) .. (60.02,10.4) .. controls (60.61,10.85) and (59.31,12.46) .. (60.02,12.07) .. controls (61.9,11.03) and (61.8,10.89) .. (63.33,9.5) .. controls (63.86,9.01) and (64.09,10.32) .. (64.43,9.95) .. controls (64.72,9.63) and (64.45,9.22) .. (64.7,8.89) .. controls (65.31,8.12) and (65.07,11.96) .. (66.91,11.16) .. controls (67.84,10.75) and (68.41,10.11) .. (68.83,9.5) .. controls (69.21,8.95) and (69.93,7.25) .. (69.93,7.84) .. controls (69.93,8.59) and (69.84,9.35) .. (69.93,10.1) .. controls (69.96,10.33) and (70.36,10.92) .. (70.49,10.71) .. controls (70.97,9.86) and (71.24,8.96) .. (71.86,8.14) .. controls (72.26,7.61) and (72.7,9.37) .. (72.96,9.05) .. controls (73.34,8.58) and (73.25,8.03) .. (73.51,7.54) .. controls (73.83,6.95) and (74.15,8.74) .. (74.89,9.2) .. controls (75.8,9.75) and (77.1,8.23) .. (77.37,8.59) .. controls (77.74,9.1) and (78.78,10.48) .. (79.02,9.95) .. controls (79.52,8.87) and (80.14,7.74) .. (79.85,6.63) .. controls (79.8,6.43) and (78.49,6.49) .. (78.75,6.63) .. controls (79.16,6.85) and (80.28,6.62) .. (80.4,6.93) .. controls (80.81,8.02) and (78.12,10.12) .. (80.12,10.25) .. controls (82.21,10.4) and (81.12,7.9) .. (82.6,7.08) .. controls (83.32,6.69) and (82.71,8.2) .. (82.88,8.74) .. controls (83.08,9.39) and (83.42,8.15) .. (83.7,8.74) .. controls (84.24,9.87) and (82.76,11.39) .. (84.26,12.22) .. controls (85.43,12.86) and (86.13,10.72) .. (87.01,9.95) .. controls (88.11,8.99) and (88.94,12.16) .. (88.94,11.01) ;
\draw  [color={rgb, 255:red, 0; green, 0; blue, 0 }  ][line width=0.75] [line join = round][line cap = round] (89.49,9.8) .. controls (90.17,9.05) and (91.68,9.83) .. (91.69,9.8) .. controls (92.14,8.36) and (91.88,6.87) .. (92.24,5.42) .. controls (92.36,4.94) and (93.75,7.9) .. (93.89,7.54) .. controls (94.36,6.35) and (93.87,5.08) .. (94.45,3.91) .. controls (94.57,3.65) and (95.56,3.38) .. (95.82,3.61) .. controls (96.64,4.34) and (96.71,6.78) .. (97.47,6.03) .. controls (98.51,5.01) and (97.26,3.7) .. (98.03,2.85) .. controls (98.17,2.69) and (98.79,2.37) .. (98.85,2.55) .. controls (99.4,4.24) and (98.69,6.06) .. (99.68,7.69) .. controls (101.16,10.12) and (101.52,6.42) .. (102.43,5.42) .. controls (103.18,4.61) and (102.66,7.25) .. (102.98,8.14) .. controls (103.2,8.74) and (104.26,9.15) .. (104.91,9.65) .. controls (105.61,10.18) and (106.26,8.56) .. (106.84,7.99) .. controls (107.37,7.46) and (106.61,9.21) .. (106.84,9.8) .. controls (107.04,10.32) and (108.84,11.54) .. (108.77,11.01) .. controls (108.66,10.29) and (108.03,9.61) .. (107.94,8.89) .. controls (107.92,8.75) and (108.46,8.45) .. (108.49,8.59) .. controls (108.61,9.2) and (108.25,9.97) .. (109.04,10.4) .. controls (109.48,10.64) and (109.53,9.78) .. (109.87,9.5) .. controls (110.42,9.05) and (111.38,10.29) .. (111.52,9.95) .. controls (111.99,8.86) and (110.9,7.42) .. (112.35,6.63) .. controls (112.65,6.46) and (113.44,9.95) .. (113.45,9.95) .. controls (114.4,9.43) and (114.9,8.7) .. (115.65,8.29) .. controls (115.78,8.22) and (116.13,8.2) .. (116.2,8.29) .. controls (116.71,8.99) and (116.82,9.78) .. (117.58,10.4) .. controls (117.92,10.68) and (118.67,10.08) .. (118.68,10.1) .. controls (119.15,10.88) and (118.76,11.81) .. (119.51,12.52) .. controls (119.81,12.81) and (121.23,8.63) .. (121.43,8.44) .. controls (121.71,8.19) and (122.21,7.39) .. (122.26,7.69) .. controls (122.44,8.79) and (121.95,9.95) .. (122.54,11.01) .. controls (122.86,11.59) and (122.62,9.75) .. (123.09,9.2) .. controls (123.45,8.77) and (123.94,10.02) .. (124.46,10.4) .. controls (125.43,11.11) and (126.89,9.11) .. (126.94,9.5) .. controls (127.05,10.3) and (125.63,11.34) .. (126.67,11.91) .. controls (128.18,12.74) and (127.68,9.29) .. (128.32,8.59) .. controls (128.62,8.26) and (128.4,9.39) .. (128.87,9.65) .. controls (129.31,9.89) and (129.54,8.42) .. (129.7,8.74) .. controls (130.11,9.59) and (129.19,11.31) .. (130.8,11.31) .. controls (133.24,11.31) and (132.53,8.81) .. (133.28,7.54) .. controls (133.43,7.28) and (133.83,8.56) .. (133.83,8.29) .. controls (133.83,7.94) and (133.65,7.57) .. (133.83,7.23) .. controls (133.98,6.94) and (134.82,9.44) .. (135.75,9.65) .. controls (137.52,10.04) and (136.4,7.64) .. (136.58,7.38) .. controls (136.73,7.18) and (137.16,7.76) .. (137.13,7.99) .. controls (137.04,8.74) and (137.14,9.51) .. (137.41,10.25) .. controls (137.5,10.5) and (137.12,10.81) .. (137.41,11.01) .. controls (138.67,11.87) and (139.99,8.8) .. (140.16,9.5) .. controls (140.38,10.4) and (139.05,12.22) .. (140.71,12.22) ;
\draw  [color={rgb, 255:red, 0; green, 0; blue, 0 }  ][line width=0.75] [line join = round][line cap = round] (88.66,10.56) .. controls (88.66,12.33) and (89.05,9.69) .. (90.31,9.35) ;

\draw  [color={rgb, 255:red, 0; green, 0; blue, 0 }  ][line width=0.75] [line join = round][line cap = round] (140.71,9.64) .. controls (140.71,9.25) and (141.86,8.36) .. (142.09,8.73) .. controls (142.76,9.84) and (142.29,11.06) .. (142.64,12.21) .. controls (142.79,12.7) and (143.47,10.77) .. (144.29,11) .. controls (145.69,11.38) and (145.53,12.59) .. (146.5,13.26) .. controls (147.54,13.99) and (147.78,11.55) .. (148.42,10.7) .. controls (149.08,9.82) and (149.68,11.42) .. (149.8,11.15) .. controls (150.14,10.41) and (149.37,9.42) .. (150.35,8.89) .. controls (151.98,7.99) and (151.02,10.19) .. (151.45,10.55) .. controls (152.03,11.03) and (153.11,9.15) .. (153.38,9.64) .. controls (153.87,10.54) and (153.76,11.59) .. (154.76,12.36) .. controls (155.6,13.01) and (156.03,9.3) .. (156.13,10.09) .. controls (156.31,11.41) and (156.2,12.77) .. (156.96,14.02) .. controls (157.47,14.87) and (157.9,10.59) .. (158.34,11.45) .. controls (158.84,12.44) and (158.6,13.55) .. (159.44,14.47) .. controls (159.64,14.7) and (160.14,15.32) .. (160.27,15.08) .. controls (161.26,13.19) and (161.75,11.23) .. (162.74,9.34) .. controls (163.12,8.63) and (163.8,11.24) .. (163.85,11.3) .. controls (164.08,11.62) and (164.86,9.74) .. (165.5,10.7) .. controls (165.94,11.36) and (164.51,12.22) .. (165.22,12.81) .. controls (166.02,13.47) and (166.96,10.49) .. (167.43,11) .. controls (168.09,11.72) and (169.53,11.8) .. (170.46,12.51) .. controls (171.04,12.96) and (169.74,14.56) .. (170.46,14.17) .. controls (172.34,13.14) and (172.24,12.99) .. (173.76,11.6) .. controls (174.3,11.11) and (174.53,12.42) .. (174.86,12.06) .. controls (175.15,11.74) and (174.88,11.33) .. (175.14,11) .. controls (175.74,10.22) and (175.51,14.07) .. (177.34,13.26) .. controls (178.27,12.85) and (178.85,12.22) .. (179.27,11.6) .. controls (179.65,11.05) and (180.37,9.35) .. (180.37,9.94) .. controls (180.37,10.7) and (180.28,11.45) .. (180.37,12.21) .. controls (180.4,12.43) and (180.8,13.03) .. (180.92,12.81) .. controls (181.41,11.96) and (181.68,11.07) .. (182.3,10.24) .. controls (182.69,9.72) and (183.14,11.47) .. (183.4,11.15) .. controls (183.78,10.68) and (183.68,10.13) .. (183.95,9.64) .. controls (184.27,9.06) and (184.58,10.85) .. (185.33,11.3) .. controls (186.23,11.85) and (187.54,10.33) .. (187.81,10.7) .. controls (188.17,11.2) and (189.22,12.58) .. (189.46,12.06) .. controls (189.95,10.97) and (190.57,9.84) .. (190.28,8.73) .. controls (190.23,8.54) and (188.92,8.59) .. (189.18,8.73) .. controls (189.59,8.96) and (190.72,8.73) .. (190.83,9.04) .. controls (191.25,10.12) and (188.55,12.22) .. (190.56,12.36) .. controls (192.64,12.5) and (191.55,10) .. (193.04,9.19) .. controls (193.75,8.79) and (193.15,10.3) .. (193.31,10.85) .. controls (193.51,11.5) and (193.85,10.25) .. (194.14,10.85) .. controls (194.68,11.97) and (193.19,13.5) .. (194.69,14.32) .. controls (195.86,14.96) and (196.57,12.83) .. (197.44,12.06) .. controls (198.55,11.09) and (199.37,14.27) .. (199.37,13.11) ;
\draw  [color={rgb, 255:red, 0; green, 0; blue, 0 }  ][line width=0.75] [line join = round][line cap = round] (199.92,11.91) .. controls (200.6,11.16) and (202.12,11.93) .. (202.13,11.91) .. controls (202.58,10.46) and (202.31,8.98) .. (202.68,7.53) .. controls (202.8,7.04) and (204.18,10.01) .. (204.33,9.64) .. controls (204.8,8.46) and (204.31,7.19) .. (204.88,6.02) .. controls (205.01,5.76) and (206,5.48) .. (206.26,5.72) .. controls (207.08,6.45) and (207.15,8.88) .. (207.91,8.13) .. controls (208.94,7.11) and (207.69,5.8) .. (208.46,4.96) .. controls (208.61,4.8) and (209.23,4.48) .. (209.29,4.66) .. controls (209.83,6.35) and (209.12,8.16) .. (210.11,9.79) .. controls (211.59,12.23) and (211.95,8.53) .. (212.87,7.53) .. controls (213.61,6.71) and (213.09,9.35) .. (213.42,10.24) .. controls (213.64,10.85) and (214.69,11.25) .. (215.34,11.75) .. controls (216.04,12.29) and (216.7,10.67) .. (217.27,10.09) .. controls (217.8,9.56) and (217.05,11.31) .. (217.27,11.91) .. controls (217.47,12.43) and (219.28,13.65) .. (219.2,13.11) .. controls (219.1,12.4) and (218.46,11.72) .. (218.37,11) .. controls (218.36,10.86) and (218.9,10.56) .. (218.93,10.7) .. controls (219.05,11.31) and (218.69,12.08) .. (219.48,12.51) .. controls (219.91,12.75) and (219.96,11.88) .. (220.3,11.6) .. controls (220.85,11.15) and (221.81,12.39) .. (221.95,12.06) .. controls (222.42,10.97) and (221.34,9.53) .. (222.78,8.73) .. controls (223.08,8.57) and (223.88,12.06) .. (223.88,12.06) .. controls (224.84,11.53) and (225.34,10.81) .. (226.09,10.4) .. controls (226.22,10.32) and (226.57,10.3) .. (226.64,10.4) .. controls (227.14,11.09) and (227.26,11.89) .. (228.01,12.51) .. controls (228.35,12.79) and (229.1,12.19) .. (229.11,12.21) .. controls (229.59,12.99) and (229.2,13.91) .. (229.94,14.62) .. controls (230.24,14.91) and (231.66,10.74) .. (231.87,10.55) .. controls (232.14,10.3) and (232.65,9.5) .. (232.7,9.79) .. controls (232.88,10.9) and (232.39,12.05) .. (232.97,13.11) .. controls (233.29,13.7) and (233.05,11.86) .. (233.52,11.3) .. controls (233.88,10.87) and (234.38,12.13) .. (234.9,12.51) .. controls (235.86,13.21) and (237.32,11.22) .. (237.38,11.6) .. controls (237.49,12.41) and (236.06,13.45) .. (237.1,14.02) .. controls (238.61,14.85) and (238.12,11.39) .. (238.75,10.7) .. controls (239.05,10.37) and (238.83,11.5) .. (239.3,11.75) .. controls (239.74,11.99) and (239.97,10.52) .. (240.13,10.85) .. controls (240.54,11.7) and (239.63,13.42) .. (241.23,13.42) .. controls (243.67,13.42) and (242.97,10.91) .. (243.71,9.64) .. controls (243.86,9.38) and (244.26,10.67) .. (244.26,10.4) .. controls (244.26,10.04) and (244.09,9.68) .. (244.26,9.34) .. controls (244.41,9.05) and (245.25,11.55) .. (246.19,11.75) .. controls (247.95,12.14) and (246.83,9.74) .. (247.02,9.49) .. controls (247.17,9.28) and (247.59,9.87) .. (247.57,10.09) .. controls (247.47,10.85) and (247.57,11.62) .. (247.84,12.36) .. controls (247.93,12.61) and (247.56,12.92) .. (247.84,13.11) .. controls (249.1,13.98) and (250.43,10.91) .. (250.6,11.6) .. controls (250.82,12.51) and (249.48,14.32) .. (251.15,14.32) ;
\draw  [color={rgb, 255:red, 0; green, 0; blue, 0 }  ][line width=0.75] [line join = round][line cap = round] (199.1,12.66) .. controls (199.1,14.43) and (199.49,11.8) .. (200.75,11.45) ;

\draw  [color={rgb, 255:red, 0; green, 0; blue, 0 }  ][line width=0.75] [line join = round][line cap = round] (30.28,66.48) .. controls (30.28,66.09) and (31.43,65.2) .. (31.65,65.58) .. controls (32.32,66.68) and (31.86,67.9) .. (32.2,69.05) .. controls (32.36,69.55) and (33.04,67.62) .. (33.86,67.84) .. controls (35.25,68.22) and (35.1,69.43) .. (36.06,70.11) .. controls (37.1,70.83) and (37.35,68.4) .. (37.99,67.54) .. controls (38.65,66.67) and (39.24,68.26) .. (39.37,67.99) .. controls (39.7,67.25) and (38.93,66.27) .. (39.92,65.73) .. controls (41.54,64.84) and (40.58,67.03) .. (41.02,67.39) .. controls (41.6,67.87) and (42.68,65.99) .. (42.95,66.48) .. controls (43.44,67.38) and (43.33,68.44) .. (44.32,69.2) .. controls (45.17,69.85) and (45.6,66.14) .. (45.7,66.94) .. controls (45.87,68.25) and (45.77,69.61) .. (46.53,70.86) .. controls (47.04,71.71) and (47.47,67.44) .. (47.9,68.29) .. controls (48.4,69.28) and (48.17,70.4) .. (49,71.31) .. controls (49.21,71.54) and (49.7,72.16) .. (49.83,71.92) .. controls (50.83,70.03) and (51.31,68.07) .. (52.31,66.18) .. controls (52.68,65.47) and (53.37,68.08) .. (53.41,68.14) .. controls (53.65,68.47) and (54.43,66.58) .. (55.06,67.54) .. controls (55.5,68.2) and (54.07,69.07) .. (54.79,69.65) .. controls (55.58,70.31) and (56.53,67.34) .. (56.99,67.84) .. controls (57.65,68.57) and (59.09,68.64) .. (60.02,69.35) .. controls (60.61,69.8) and (59.31,71.4) .. (60.02,71.01) .. controls (61.9,69.98) and (61.8,69.84) .. (63.33,68.45) .. controls (63.86,67.96) and (64.09,69.26) .. (64.43,68.9) .. controls (64.72,68.58) and (64.45,68.17) .. (64.7,67.84) .. controls (65.31,67.07) and (65.07,70.91) .. (66.91,70.11) .. controls (67.84,69.7) and (68.41,69.06) .. (68.83,68.45) .. controls (69.21,67.89) and (69.93,66.2) .. (69.93,66.78) .. controls (69.93,67.54) and (69.84,68.3) .. (69.93,69.05) .. controls (69.96,69.27) and (70.36,69.87) .. (70.49,69.65) .. controls (70.97,68.8) and (71.24,67.91) .. (71.86,67.09) .. controls (72.26,66.56) and (72.7,68.31) .. (72.96,67.99) .. controls (73.34,67.52) and (73.25,66.97) .. (73.51,66.48) .. controls (73.83,65.9) and (74.15,67.69) .. (74.89,68.14) .. controls (75.8,68.7) and (77.1,67.18) .. (77.37,67.54) .. controls (77.74,68.05) and (78.78,69.43) .. (79.02,68.9) .. controls (79.52,67.81) and (80.14,66.68) .. (79.85,65.58) .. controls (79.8,65.38) and (78.49,65.43) .. (78.75,65.58) .. controls (79.16,65.8) and (80.28,65.57) .. (80.4,65.88) .. controls (80.81,66.96) and (78.12,69.06) .. (80.12,69.2) .. controls (82.21,69.34) and (81.12,66.84) .. (82.6,66.03) .. controls (83.32,65.64) and (82.71,67.14) .. (82.88,67.69) .. controls (83.08,68.34) and (83.42,67.09) .. (83.7,67.69) .. controls (84.24,68.81) and (82.76,70.34) .. (84.26,71.16) .. controls (85.43,71.81) and (86.13,69.67) .. (87.01,68.9) .. controls (88.11,67.93) and (88.94,71.11) .. (88.94,69.96) ;
\draw  [color={rgb, 255:red, 0; green, 0; blue, 0 }  ][line width=0.75] [line join = round][line cap = round] (89.49,68.75) .. controls (90.17,68) and (91.68,68.78) .. (91.69,68.75) .. controls (92.14,67.31) and (91.88,65.82) .. (92.24,64.37) .. controls (92.36,63.89) and (93.75,66.85) .. (93.89,66.48) .. controls (94.36,65.3) and (93.87,64.03) .. (94.45,62.86) .. controls (94.57,62.6) and (95.56,62.33) .. (95.82,62.56) .. controls (96.64,63.29) and (96.71,65.72) .. (97.47,64.97) .. controls (98.51,63.95) and (97.26,62.65) .. (98.03,61.8) .. controls (98.17,61.64) and (98.79,61.32) .. (98.85,61.5) .. controls (99.4,63.19) and (98.69,65) .. (99.68,66.63) .. controls (101.16,69.07) and (101.52,65.37) .. (102.43,64.37) .. controls (103.18,63.55) and (102.66,66.19) .. (102.98,67.09) .. controls (103.2,67.69) and (104.26,68.1) .. (104.91,68.6) .. controls (105.61,69.13) and (106.26,67.51) .. (106.84,66.94) .. controls (107.37,66.41) and (106.61,68.16) .. (106.84,68.75) .. controls (107.04,69.27) and (108.84,70.49) .. (108.77,69.96) .. controls (108.66,69.24) and (108.03,68.56) .. (107.94,67.84) .. controls (107.92,67.7) and (108.46,67.4) .. (108.49,67.54) .. controls (108.61,68.15) and (108.25,68.92) .. (109.04,69.35) .. controls (109.48,69.59) and (109.53,68.73) .. (109.87,68.45) .. controls (110.42,68) and (111.38,69.24) .. (111.52,68.9) .. controls (111.99,67.81) and (110.9,66.37) .. (112.35,65.58) .. controls (112.65,65.41) and (113.44,68.9) .. (113.45,68.9) .. controls (114.4,68.38) and (114.9,67.65) .. (115.65,67.24) .. controls (115.78,67.17) and (116.13,67.14) .. (116.2,67.24) .. controls (116.71,67.93) and (116.82,68.73) .. (117.58,69.35) .. controls (117.92,69.63) and (118.67,69.03) .. (118.68,69.05) .. controls (119.15,69.83) and (118.76,70.75) .. (119.51,71.47) .. controls (119.81,71.75) and (121.23,67.58) .. (121.43,67.39) .. controls (121.71,67.14) and (122.21,66.34) .. (122.26,66.63) .. controls (122.44,67.74) and (121.95,68.89) .. (122.54,69.96) .. controls (122.86,70.54) and (122.62,68.7) .. (123.09,68.14) .. controls (123.45,67.71) and (123.94,68.97) .. (124.46,69.35) .. controls (125.43,70.06) and (126.89,68.06) .. (126.94,68.45) .. controls (127.05,69.25) and (125.63,70.29) .. (126.67,70.86) .. controls (128.18,71.69) and (127.68,68.24) .. (128.32,67.54) .. controls (128.62,67.21) and (128.4,68.34) .. (128.87,68.6) .. controls (129.31,68.84) and (129.54,67.36) .. (129.7,67.69) .. controls (130.11,68.54) and (129.19,70.26) .. (130.8,70.26) .. controls (133.24,70.26) and (132.53,67.76) .. (133.28,66.48) .. controls (133.43,66.22) and (133.83,67.51) .. (133.83,67.24) .. controls (133.83,66.89) and (133.65,66.52) .. (133.83,66.18) .. controls (133.98,65.89) and (134.82,68.39) .. (135.75,68.6) .. controls (137.52,68.98) and (136.4,66.59) .. (136.58,66.33) .. controls (136.73,66.12) and (137.16,66.71) .. (137.13,66.94) .. controls (137.04,67.69) and (137.14,68.46) .. (137.41,69.2) .. controls (137.5,69.45) and (137.12,69.76) .. (137.41,69.96) .. controls (138.67,70.82) and (139.99,67.75) .. (140.16,68.45) .. controls (140.38,69.35) and (139.05,71.16) .. (140.71,71.16) ;
\draw  [color={rgb, 255:red, 0; green, 0; blue, 0 }  ][line width=0.75] [line join = round][line cap = round] (88.66,69.5) .. controls (88.66,71.27) and (89.05,68.64) .. (90.31,68.29) ;

\draw  [color={rgb, 255:red, 0; green, 0; blue, 0 }  ][line width=0.75] [line join = round][line cap = round] (140.71,68.59) .. controls (140.71,68.2) and (141.86,67.31) .. (142.09,67.68) .. controls (142.76,68.78) and (142.29,70.01) .. (142.64,71.16) .. controls (142.79,71.65) and (143.47,69.72) .. (144.29,69.95) .. controls (145.69,70.33) and (145.53,71.54) .. (146.5,72.21) .. controls (147.54,72.94) and (147.78,70.5) .. (148.42,69.65) .. controls (149.08,68.77) and (149.68,70.36) .. (149.8,70.1) .. controls (150.14,69.36) and (149.37,68.37) .. (150.35,67.83) .. controls (151.98,66.94) and (151.02,69.14) .. (151.45,69.49) .. controls (152.03,69.97) and (153.11,68.1) .. (153.38,68.59) .. controls (153.87,69.49) and (153.76,70.54) .. (154.76,71.31) .. controls (155.6,71.95) and (156.03,68.25) .. (156.13,69.04) .. controls (156.31,70.36) and (156.2,71.72) .. (156.96,72.97) .. controls (157.47,73.81) and (157.9,69.54) .. (158.34,70.4) .. controls (158.84,71.39) and (158.6,72.5) .. (159.44,73.42) .. controls (159.64,73.65) and (160.14,74.27) .. (160.27,74.02) .. controls (161.26,72.14) and (161.75,70.17) .. (162.74,68.29) .. controls (163.12,67.58) and (163.8,70.19) .. (163.85,70.25) .. controls (164.08,70.57) and (164.86,68.68) .. (165.5,69.65) .. controls (165.94,70.31) and (164.51,71.17) .. (165.22,71.76) .. controls (166.02,72.41) and (166.96,69.44) .. (167.43,69.95) .. controls (168.09,70.67) and (169.53,70.74) .. (170.46,71.46) .. controls (171.04,71.91) and (169.74,73.51) .. (170.46,73.12) .. controls (172.34,72.09) and (172.24,71.94) .. (173.76,70.55) .. controls (174.3,70.06) and (174.53,71.37) .. (174.86,71) .. controls (175.15,70.69) and (174.88,70.27) .. (175.14,69.95) .. controls (175.74,69.17) and (175.51,73.02) .. (177.34,72.21) .. controls (178.27,71.8) and (178.85,71.17) .. (179.27,70.55) .. controls (179.65,70) and (180.37,68.3) .. (180.37,68.89) .. controls (180.37,69.65) and (180.28,70.4) .. (180.37,71.16) .. controls (180.4,71.38) and (180.8,71.97) .. (180.92,71.76) .. controls (181.41,70.91) and (181.68,70.02) .. (182.3,69.19) .. controls (182.69,68.66) and (183.14,70.42) .. (183.4,70.1) .. controls (183.78,69.63) and (183.68,69.08) .. (183.95,68.59) .. controls (184.27,68.01) and (184.58,69.8) .. (185.33,70.25) .. controls (186.23,70.8) and (187.54,69.28) .. (187.81,69.65) .. controls (188.17,70.15) and (189.22,71.53) .. (189.46,71) .. controls (189.95,69.92) and (190.57,68.79) .. (190.28,67.68) .. controls (190.23,67.48) and (188.92,67.54) .. (189.18,67.68) .. controls (189.59,67.91) and (190.72,67.67) .. (190.83,67.98) .. controls (191.25,69.07) and (188.55,71.17) .. (190.56,71.31) .. controls (192.64,71.45) and (191.55,68.95) .. (193.04,68.14) .. controls (193.75,67.74) and (193.15,69.25) .. (193.31,69.8) .. controls (193.51,70.45) and (193.85,69.2) .. (194.14,69.8) .. controls (194.68,70.92) and (193.19,72.45) .. (194.69,73.27) .. controls (195.86,73.91) and (196.57,71.77) .. (197.44,71) .. controls (198.55,70.04) and (199.37,73.21) .. (199.37,72.06) ;
\draw  [color={rgb, 255:red, 0; green, 0; blue, 0 }  ][line width=0.75] [line join = round][line cap = round] (199.92,70.85) .. controls (200.6,70.11) and (202.12,70.88) .. (202.13,70.85) .. controls (202.58,69.41) and (202.31,67.92) .. (202.68,66.47) .. controls (202.8,65.99) and (204.18,68.96) .. (204.33,68.59) .. controls (204.8,67.4) and (204.31,66.13) .. (204.88,64.96) .. controls (205.01,64.7) and (206,64.43) .. (206.26,64.66) .. controls (207.08,65.39) and (207.15,67.83) .. (207.91,67.08) .. controls (208.94,66.06) and (207.69,64.75) .. (208.46,63.91) .. controls (208.61,63.75) and (209.23,63.43) .. (209.29,63.61) .. controls (209.83,65.3) and (209.12,67.11) .. (210.11,68.74) .. controls (211.59,71.17) and (211.95,67.48) .. (212.87,66.47) .. controls (213.61,65.66) and (213.09,68.3) .. (213.42,69.19) .. controls (213.64,69.79) and (214.69,70.2) .. (215.34,70.7) .. controls (216.04,71.24) and (216.7,69.62) .. (217.27,69.04) .. controls (217.8,68.51) and (217.05,70.26) .. (217.27,70.85) .. controls (217.47,71.38) and (219.28,72.59) .. (219.2,72.06) .. controls (219.1,71.34) and (218.46,70.67) .. (218.37,69.95) .. controls (218.36,69.81) and (218.9,69.5) .. (218.93,69.65) .. controls (219.05,70.25) and (218.69,71.02) .. (219.48,71.46) .. controls (219.91,71.7) and (219.96,70.83) .. (220.3,70.55) .. controls (220.85,70.1) and (221.81,71.34) .. (221.95,71) .. controls (222.42,69.92) and (221.34,68.47) .. (222.78,67.68) .. controls (223.08,67.52) and (223.88,71.01) .. (223.88,71) .. controls (224.84,70.48) and (225.34,69.75) .. (226.09,69.34) .. controls (226.22,69.27) and (226.57,69.25) .. (226.64,69.34) .. controls (227.14,70.04) and (227.26,70.83) .. (228.01,71.46) .. controls (228.35,71.73) and (229.1,71.14) .. (229.11,71.16) .. controls (229.59,71.93) and (229.2,72.86) .. (229.94,73.57) .. controls (230.24,73.86) and (231.66,69.68) .. (231.87,69.49) .. controls (232.14,69.24) and (232.65,68.45) .. (232.7,68.74) .. controls (232.88,69.84) and (232.39,71) .. (232.97,72.06) .. controls (233.29,72.65) and (233.05,70.81) .. (233.52,70.25) .. controls (233.88,69.82) and (234.38,71.08) .. (234.9,71.46) .. controls (235.86,72.16) and (237.32,70.17) .. (237.38,70.55) .. controls (237.49,71.36) and (236.06,72.4) .. (237.1,72.97) .. controls (238.61,73.8) and (238.12,70.34) .. (238.75,69.65) .. controls (239.05,69.32) and (238.83,70.44) .. (239.3,70.7) .. controls (239.74,70.94) and (239.97,69.47) .. (240.13,69.8) .. controls (240.54,70.65) and (239.63,72.36) .. (241.23,72.36) .. controls (243.67,72.36) and (242.97,69.86) .. (243.71,68.59) .. controls (243.86,68.33) and (244.26,69.61) .. (244.26,69.34) .. controls (244.26,68.99) and (244.09,68.63) .. (244.26,68.29) .. controls (244.41,67.99) and (245.25,70.5) .. (246.19,70.7) .. controls (247.95,71.09) and (246.83,68.69) .. (247.02,68.44) .. controls (247.17,68.23) and (247.59,68.82) .. (247.57,69.04) .. controls (247.47,69.8) and (247.57,70.56) .. (247.84,71.31) .. controls (247.93,71.55) and (247.56,71.86) .. (247.84,72.06) .. controls (249.1,72.92) and (250.43,69.86) .. (250.6,70.55) .. controls (250.82,71.45) and (249.48,73.27) .. (251.15,73.27) ;
\draw  [color={rgb, 255:red, 0; green, 0; blue, 0 }  ][line width=0.75] [line join = round][line cap = round] (199.1,71.61) .. controls (199.1,73.38) and (199.49,70.75) .. (200.75,70.4) ;

\draw  [draw opacity=0][fill={rgb, 255:red, 255; green, 255; blue, 255 }  ,fill opacity=1 ] (154.52,0.39) -- (168.32,0.39) -- (168.32,84.61) -- (154.52,84.61) -- cycle ;
\draw [color={rgb, 255:red, 144; green, 19; blue, 254 }  ,draw opacity=1 ] [dash pattern={on 4.5pt off 4.5pt}]  (154.52,0.39) -- (154.52,84.61) ;
\draw [color={rgb, 255:red, 144; green, 19; blue, 254 }  ,draw opacity=1 ] [dash pattern={on 4.5pt off 4.5pt}]  (168.32,0.39) -- (168.32,84.61) ;
\draw   (30.28,0.39) -- (251.15,0.39) -- (251.15,84.61) -- (30.28,84.61) -- cycle ;
\draw [color={rgb, 255:red, 144; green, 19; blue, 254 }  ,draw opacity=1 ]   (164.42,42.5) -- (195.93,42.5) ;
\draw [shift={(161.42,42.5)}, rotate = 0] [fill={rgb, 255:red, 144; green, 19; blue, 254 }  ,fill opacity=1 ][line width=0.08]  [draw opacity=0] (8.93,-4.29) -- (0,0) -- (8.93,4.29) -- cycle    ;
\draw [color={rgb, 255:red, 208; green, 2; blue, 27 }  ,draw opacity=1 ]   (127.6,31.8) -- (99.3,53.03) ;
\draw [shift={(130,30)}, rotate = 143.13] [fill={rgb, 255:red, 208; green, 2; blue, 27 }  ,fill opacity=1 ][line width=0.08]  [draw opacity=0] (8.93,-4.29) -- (0,0) -- (8.93,4.29) -- cycle    ;
\draw [color={rgb, 255:red, 144; green, 19; blue, 254 }  ,draw opacity=1 ][line width=1.5]    (154.25,70.97) -- (168.69,71) ;
\draw [color={rgb, 255:red, 144; green, 19; blue, 254 }  ,draw opacity=1 ][line width=1.5]    (154.25,22.97) -- (168.69,23) ;
\draw [color={rgb, 255:red, 144; green, 19; blue, 254 }  ,draw opacity=1 ][line width=1.5]    (154.25,11.74) -- (168.69,11.77) ;

\draw (126.15,43.79) node [anchor=north west][inner sep=0.75pt]  [font=\small]  {$\dotsc $};
\draw (66.44,87) node [anchor=north west][inner sep=0.75pt]  [font=\scriptsize] [align=left] {$\displaystyle n$ time samples};
\draw (2,68.29) node [anchor=north west][inner sep=0.75pt]  [font=\scriptsize,rotate=-270] [align=left] {$\displaystyle p$ electrodes};
\draw (204.54,34.74) node [anchor=north west][inner sep=0.75pt]  [font=\scriptsize,color={rgb, 255:red, 144; green, 19; blue, 254 }  ,opacity=1 ] [align=left] {\begin{minipage}[lt]{26.93pt}\setlength\topsep{0pt}
\begin{center}
eye \\blinking
\end{center}

\end{minipage}};
\draw (45.74,36.84) node [anchor=north west][inner sep=0.75pt]  [font=\scriptsize,color={rgb, 255:red, 208; green, 2; blue, 27 }  ,opacity=1 ] [align=left] {\begin{minipage}[lt]{32.09pt}\setlength\topsep{0pt}
\begin{center}
electrode\\popping
\end{center}

\end{minipage}};

\end{tikzpicture}

%% file: img/popping.tex
\begin{tikzpicture}

\definecolor{darkslateblue5793155}{RGB}{57,93,155}
\definecolor{darkslategray38}{RGB}{38,38,38}
\definecolor{darkslategray564284}{RGB}{56,42,84}
\definecolor{lavender234234242}{RGB}{234,234,242}
\definecolor{lightgray204}{RGB}{204,204,204}
\definecolor{mediumaquamarine95206172}{RGB}{95,206,172}
\definecolor{steelblue52150169}{RGB}{52,150,169}
\definecolor{redpastel}{RGB}{255,38,38}
\definecolor{orangepastel}{RGB}{255,157,46}
\definecolor{bluepastel}{RGB}{46,56,169}
\definecolor{lightblue}{RGB}{128, 168, 255}

\begin{axis}[at={(.47\linewidth,0)},width=5cm,height=4cm,
title={P300},
title style={at={(.5,.9)}},
axis background/.style={fill=lavender234234242},
axis line style={white},
legend cell align={left},
legend style={
  text opacity=1,
  at={(-0.01,0)},
  anchor=south west,
  fill=none,
  draw=none,
  font={\footnotesize}
},
tick align=outside,
x grid style={white},
xlabel=\textcolor{darkslategray38}{Incomplete epochs (\%)},
xmin=-0.5, xmax=3.5,
xtick style={color=darkslategray38},
xtick={0,1,2,3},
xticklabels={10,20,30,40},
y grid style={white},
ymajorgrids,
ymin=78.5105545782023, ymax=88.3822828181285,
ytick style={color=darkslategray38,
font={\scriptsize}},
ytick={78,80,82,84,86,88,90},
yticklabels={
  \(\displaystyle {78}\),
  \(\displaystyle {80}\),
  \(\displaystyle {82}\),
  \(\displaystyle {84}\),
  \(\displaystyle {86}\),
  \(\displaystyle {88}\),
  \(\displaystyle {90}\)
},
xtick pos=left,
ytick pos=left
]
\addplot [line width=1.08pt, bluepastel, forget plot]
table {%
0 87.9335708618164
3 87.9335708618164
};
\addplot [draw=bluepastel, fill=bluepastel, mark=triangle*, mark size=3pt, only marks]
table{%
x  y
0 87.9335678981318
1 87.9335678981318
2 87.9335678981318
3 87.9335678981318
};
\addlegendentry{SCM}
\addplot [line width=1.08pt, redpastel, forget plot]
table {%
0 86.3770294189453
1 84.9184417724609
2 81.6109924316406
3 81.1008987426758
};
\addplot [mark=*, draw=redpastel, fill=redpastel, mark size=2.5pt, only marks]
table{%
x  y
0 86.3770294043729
1 84.918438468627
2 81.6109910362738
3 81.1008963726229
};
\addlegendentry{EM-SCM}
\addplot [line width=1.08pt, orangepastel, forget plot]
table {%
0 86.2388534545898
1 85.0232543945312
2 80.8882446289062
3 78.9592666625977
};
\addplot [draw=orangepastel, fill=orangepastel, mark=diamond*, mark size=3pt, only marks]
table{%
x  y
0 86.2388539834129
1 85.023255424311
2 80.888246628131
3 78.9592694981989
};
\addlegendentry{Mask-SCM}
\end{axis}
\begin{axis}[at={(0,0)},height=4cm,width=5cm,
title={Motor Imagery},
title style={at={(0.5,.88)}},
axis background/.style={fill=lavender234234242},
axis line style={white},
tick align=outside,
x grid style={white},
xlabel=\textcolor{white!15!black}{Incomplete epochs (\%)},
xmin=-0.5, xmax=3.5,
xtick style={color=white!15!black},
xtick={0,1,2,3},
xticklabels={10,20,30,40},
y grid style={white},
ylabel=\textcolor{white!15!black}{Mean accuracy (\%)},
ymajorgrids,
ymin=40.6386306846577, ymax=63.7245210727969,
ytick style={color=white!15!black},
ytick={42,46,50,54,58,62},
yticklabels={
	\(\displaystyle {42}\),
	\(\displaystyle {46}\),
	\(\displaystyle {50}\),
	\(\displaystyle {54}\),
	\(\displaystyle {58}\),
	\(\displaystyle {62}\)
},
ytick pos=left,
xtick pos=left,
]
\addplot [line width=1.08pt, bluepastel, forget plot]
table {%
	0 62.675163269043
	3 62.675163269043
};
\addplot [draw=bluepastel, fill=bluepastel, mark=triangle*, mark size=3pt, only marks]
table{%
	x  y
	0 62.6751624187906
	1 62.6751624187906
	2 62.6751624187906
	3 62.6751624187906
};
\addplot [line width=1.08pt, redpastel, forget plot]
table {%
	0 59.1694145202637
	1 55.9643516540527
	2 55.0398139953613
	3 45.612361907959
};
\addplot [draw=redpastel, fill=redpastel, mark=*, mark size=2.5pt, only marks]
table{%
	x  y
	0 59.1694152923538
	1 55.9643511577545
	2 55.03981342662
	3 45.6123604864235
};
\addplot [line width=1.08pt, orangepastel, forget plot]
table {%
	0 48.3744812011719
	1 48.9443626403809
	2 45.5359001159668
	3 41.68798828125
};
\addplot [draw=orangepastel, fill=orangepastel, mark=diamond*, mark size=3pt, only marks]
table{%
	x  y
	0 48.3744794269532
	1 48.944361152757
	2 45.535898717308
	3 41.687989338664
};
\end{axis}

\end{tikzpicture}

%% file: img/eye_blinking.tex
\begin{tikzpicture}

\definecolor{darkslateblue5793155}{RGB}{57,93,155}
\definecolor{darkslategray38}{RGB}{38,38,38}
\definecolor{darkslategray564284}{RGB}{56,42,84}
\definecolor{lavender234234242}{RGB}{234,234,242}
\definecolor{lightgray204}{RGB}{204,204,204}
\definecolor{mediumaquamarine95206172}{RGB}{95,206,172}
\definecolor{steelblue52150169}{RGB}{52,150,169}
\definecolor{redpastel}{RGB}{255,38,38}
\definecolor{orangepastel}{RGB}{255,157,46}
\definecolor{bluepastel}{RGB}{46,56,169}
\definecolor{lightblue}{RGB}{128, 168, 255}

\begin{axis}[at={(0,0)},width=5cm,height=4cm,
axis background/.style={fill=lavender234234242},
axis line style={white},
title={Motor Imagery},
title style={at={(.5,.88)}},
tick align=outside,
x grid style={white},
xlabel=\textcolor{darkslategray38}{Incomplete epochs (\%)},
xmin=-0.5, xmax=3.5,
xtick style={color=darkslategray38},
xtick={0,1,2,3},
xticklabels={10,20,30,40},
y grid style={white},
ylabel=\textcolor{darkslategray38}{Mean accuracy (\%)},
ymajorgrids,
ymin=52.6576711644178, ymax=63.1,
ytick style={color=darkslategray38},
ytick={53,56,59,62,65},
yticklabels={
  \(\displaystyle {53}\),
  \(\displaystyle {56}\),
  \(\displaystyle {59}\),
  \(\displaystyle {62}\),
  \(\displaystyle {65}\)
},
xtick pos=left,
ytick pos=left
]
\addplot [line width=1.08pt, bluepastel, forget plot]
table {%
0 62.675163269043
3 62.675163269043
};
\addplot [draw=bluepastel, fill=bluepastel, mark=triangle*, mark size=3pt, only marks]
table{%
x  y
0 62.6751624187906
1 62.6751624187906
2 62.6751624187906
3 62.6751624187906
};
\addplot [line width=1.08pt, redpastel, forget plot]
table {%
0 61.3050155639648
1 60.0334815979004
2 58.8737297058105
3 57.0026664733887
};
\addplot [draw=redpastel, fill=redpastel, mark=*, mark size=2.5pt, only marks]
table{%
x  y
0 61.3050141595869
1 60.0334832583708
2 58.8737298017658
3 57.0026653339997
};
\addplot [line width=1.08pt, orangepastel, forget plot]
table {%
0 60.1652488708496
1 57.4834251403809
2 55.1107788085938
3 53.220890045166
};
\addplot [draw=orangepastel, fill=orangepastel, mark=diamond*, mark size=3pt, only marks]
table{%
x  y
0 60.1652507079793
1 57.4834249541896
2 55.1107779443612
3 53.2208895552224
};
\end{axis}
\begin{axis}[at={(.47\linewidth,0)},width=5cm,height=4cm,
axis background/.style={fill=lavender234234242},
axis line style={white},
title={P300},
title style={at={(.5,.9)}},
legend cell align={left},
legend style={
	text opacity=1,
	at={(0,-0.02)},
	anchor=south west,
	draw=none,
	fill=none,
	font={\footnotesize},
},
tick align=outside,
x grid style={white},
xlabel=\textcolor{darkslategray38}{Incomplete epochs (\%)},
xmin=-0.5, xmax=3.5,
xtick style={color=darkslategray38},
xtick={0,1,2,3},
xticklabels={10,20,30,40},
y grid style={white},
ymajorgrids,
ymin=82, ymax=88.2,
ytick style={color=darkslategray38},
ytick={82,84,86,88,90},
yticklabels={
	\(\displaystyle {82}\),
	\(\displaystyle {84}\),
	\(\displaystyle {86}\),
	\(\displaystyle {88}\),
	\(\displaystyle {90}\)
},
xtick pos=left,
ytick pos=left
]
\addplot [line width=1.08pt, bluepastel, forget plot]
table {%
	0 87.9335708618164
	3 87.9335708618164
};
\addplot [draw=bluepastel, fill=bluepastel, mark=triangle*, mark size=3pt, only marks]
table{%
	x  y
	0 87.9335678981318
	1 87.9335678981318
	2 87.9335678981318
	3 87.9335678981318
};
\addlegendentry{SCM}
\addplot [line width=1.08pt, redpastel, forget plot]
table {%
	0 87.1007766723633
	1 85.561149597168
	2 82.8410491943359
	3 82.5129623413086
};
\addplot [draw=redpastel, fill=redpastel, mark=*, mark size=2.5pt, only marks]
table{%
	x  y
	0 87.1007790902237
	1 85.5611460165871
	2 82.841048839742
	3 82.5129597051185
};
\addlegendentry{EM-SCM}
\addplot [line width=1.08pt, orangepastel, forget plot]
table {%
	0 87.0315322875977
	1 85.4339752197266
	2 84.1840362548828
	3 83.1663208007812
};
\addplot [draw=orangepastel, fill=orangepastel, mark=diamond*, mark size=3pt, only marks]
table{%
	x  y
	0 87.0315322107732
	1 85.4339783865293
	2 84.1840328390718
	3 83.1663231967831
};
\addlegendentry{k-NN + SCM}
\end{axis}

\end{tikzpicture}